\documentclass[twocolumn,preprintnumbers,superscriptaddress,floatfix]{revtex4}
\usepackage{amsmath}
\usepackage{amssymb}
\usepackage{mathrsfs}
\usepackage{graphicx}
\usepackage{bm}
\usepackage{epstopdf}
\usepackage[colorlinks=true,citecolor=blue,linkcolor=blue,urlcolor=blue]{hyperref}
\usepackage{bookmark}
\usepackage{color}
\begin{document}
\title{Anisotropic and tunable optical conductivity of a two-dimensional semi-Dirac system
in the presence of elliptically polarized radiation}

\author{H. Y. Zhang}
\affiliation{School of Physics and Astronomy and Yunnan Key laboratory of Quantum Information,
Yunnan University, Kunming 650091, China}

\author{Y. M. Xiao}\email{yiming.xiao@ynu.edu.cn}
\affiliation{School of Physics and Astronomy and Yunnan Key laboratory of Quantum Information,
Yunnan University, Kunming 650091, China}

\author{Q. N. Li}
\affiliation{School of Physics and Astronomy and Yunnan Key laboratory of Quantum Information,
Yunnan University, Kunming 650091, China}

\author{L. Ding}
\affiliation{School of Physics and Astronomy and Yunnan Key laboratory of Quantum Information,
Yunnan University, Kunming 650091, China}

\author{B. Van Duppen}
\affiliation{Department of Physics, University of Antwerp, Groenenborgerlaan 171, B-2020
Antwerpen, Belgium}

\author{W. Xu}\email{wenxu$_$issp@aliyun.com}
\affiliation{Micro Optical Instruments Inc., 518118 Shenzhen, China}
\affiliation{School of Physics and Astronomy and Yunnan Key laboratory of Quantum Information,
Yunnan University, Kunming 650091, China}
\affiliation{Key Laboratory of Materials Physics, Institute of Solid State Physics, HFIPS,
Chinese Academy of Sciences, Hefei 230031, China}

\author{F. M. Peeters}
\affiliation{School of Physics and Astronomy and Yunnan Key laboratory of Quantum Information,
Yunnan University, Kunming 650091, China}
\affiliation{Department of Physics, University of Antwerp,
Groenenborgerlaan 171, B-2020 Antwerpen, Belgium}

\date{\today}
\begin{abstract}
We investigate the effect of ellipticity ratio of the polarized radiation field on
optoelectronic properties of a two-dimensional (2D) semi-Dirac (SD) system. The optical
conductivity is calculated within the energy balance equation approach derived from
the semiclassical Boltzmann equation. We find that there exists the anisotropic optical
absorption induced via both the intra- and interband electronic transition channels
in the perpendicular $xx$ and $yy$ directions. Furthermore, we examine the effects of
the ellipticity ratio, the temperature, the carrier density, and the band-gap parameter
on the optical conductivity of the 2D SD system placed in transverse and vertical
directions, respectively. It is shown that the ellipticity ratio, temperature, carrier
density, and band-gap parameter can play the important roles in tuning the strength, peak
position, and shape of the optical conductivity spectrum. The results obtained from this
study indicate that the 2D SD system can be a promising anisotropic and tunable optical
and optoelectronic material for applications in innovative 2D optical and optoelectronic
devices, which are active in the infrared and terahertz bandwidths.
\end{abstract}

\maketitle

\section{Introduction}
The isolation of a single layer of graphite, known as graphene exhibits an unique
Dirac-like linear and gapless band structure in low-energy regime, has emerged as
an important and promising field of research in condensed matter physics and
electronics since 2004 \cite{Novoselov04}. The rise of newly developed two-dimensional
(2D) electronic materials such as silicene, monolayer (ML) transition metal dichalcogenides
(TMDs), ML hexagonal boron nitride ($h$-BN), ML black phosphorus (BP), etc., are
expected to lead to applications in next generation of high-performance nano-electronic,
optical and optoelectronic devices \cite{Li17}. Recently, a distinct class of 2D materials
named 2D semi-Dirac (SD) electronic systems have been realized in materials or device
systems such as TiO$_2$/V$_2$O$_3$ superlattices \cite{Pardo09,Banerjee09}, phosphorene
under pressure \cite{Rodin14}, doped \cite{Guan14} or electrically modulated
systems \cite{Rudenko15,Dutreix16}, $\alpha$-(BEDT-TTF)$_2$I$_{3}$ salts under
pressure \cite{Kobayashi11,Suzumura13}, etc. Notably, SD fermions have an anisotropic
band structure dispersion which displays a linear dispersion along one direction and a
quadratic dispersion along the perpendicular direction \cite{Pyatkovskiy16,Mawrie19R}.
This anisotropy in the electronic band structure results in unusual physical properties
as compared to, e.g., graphene and TMD-based 2D electron gas systems. This has drawn
the attention of the scientific research community in recent years.\par

Several investigations of the electronic and optical properties of 2D SD systems have
been carried out recently. In particular, the electronic, optoelectronic,
and transport properties such as Floquet band structure \cite{Chen18,Islam18}, optical
conductivity \cite{Mawrie19,Carbotte19,Carbotte19s,Sanderson18}, dielectric function
and plasmons \cite{Pyatkovskiy16}, nonlinear response \cite{Dai19}, and ballistic
transport modulated by magnetic and electrical barriers \cite{Zhu21} have been studied.
It has been shown that the 2D SD system can transform from normal insulator to the Chern
insulating phase under light irradiation with relatively high frequencies \cite{Saha16,Sengupta21}.
Moreover, in the presence of a high magnetic field, the 2D SD system can exhibit
interesting features, such as an unusual magnetic field dependence of the Landau levels
$[(N+1/2)B]^{2/3}$ \cite{Dietl08}, the Hofstadter's butterfly \cite{Delplace10,Sinha20},
Landau-Zener oscillations \cite{Saha17}, etc. Theoretical results on the thermoelectric
properties, e.g., the Seebeck coefficient, the Ettingshausen
coefficient, and the Righi-Leduc effect suggest that 2D SD systems can also be
utilized for device applications in quantum thermoelectrics \cite{Mawrie19R,Sengupta21}.

Experimentally, SD electrons in strained black phosphorus (BP) thin films have been
realized \cite{Makino17}. The zero mode of the Landau level expected at the Dirac point
has been successfully observed in magneto-resistance measurements of
$\alpha$-(BEDT-TTF)$_2$I$_{3}$ under high hydrostatic pressure \cite{Tajima09}.
Theoretical studies found a relatively large anisotropy between the optical conductivities $\sigma^{\mathrm{inter}}_{xx}(\omega)$ and $\sigma^{\mathrm{inter}}_{yy}(\omega)$ in the
2D plane of the system \cite{Carbotte19,Mawrie19}. The polarization function has also
been evaluated. Within the random phase approximation (RPA), it was found that there
exists an undamped anisotropic collective mode with square-root dispersion
relation \cite{Banerjee12,Pyatkovskiy16}.
On the basis of the unique anisotropic band structure and of the optical properties reported,
one can predict that the polarization of the radiation light should affect the optical
response of the 2D SD system. It was shown that the 2D SD system realized from few-layer
black phosphorus at critical surface doping with potassium can exhibit linear and
quadratic bands along the armchair and zigzag directions, respectively \cite{Kim15,Baik15}.
Thus, this is an easy way to probe the anisotropic optical properties of the 2D SD system
through a geometrical device setting.

From the viewpoint of physics, the optical conductivity is a key physical quantity, which
relates directly to measurable optical coefficients such as transmittance and/or
reflectivity \cite{Mak08}. On the other hand, through optical transmission
or reflection experiments we can measure the optical conductivity in a 2D SD system under
different experimental conditions by varying the ellipticity ratio of the polarized radiation
light, temperature, carrier density, etc. In this theoretical study, we consider a 2D SD
system placed in an anisotropic geometry along different directions. The optical conductivity
or absorption under elliptically polarized light radiation with different ellipticity ratios is examined.
We find that the optical absorption differs when the sample position is rotated by $90^\mathrm{o}$
degrees. The ellipticity ratio, temperature, carrier density and band-gap parameter can effectively
tune the optical absorption in the 2D SD system, especially in terahertz (THz, 1 THz $\simeq$ 4.1 meV)
and infrared regime. Thus, from a basic physics and a device application points of view, the
results from this study show that the 2D SD system can be employed for the investigating of novel
optoelectronic physics with corresponding applications in innovative 2D optical and optoelectronic
devices.

The present paper is organized as follows. In Sec. \ref{sec:theoretical approach},
we evaluate the optical conductivity of the 2D SD system placed in different directions in the
framework of the energy balance equation derived from the semiclassical Boltzmann equation in
the presence of an elliptically polarized radiation field. The optical transition channels for
different doping levels are considered and the optical conductivities obtained under different
experimental conditions are presented and discussed in Sec. \ref{sec:results}. Our main
conclusions from this study are summarized in Sec. \ref{sec:conclusions}.

\section{Theoretical Framework}
\label{sec:theoretical approach}

In this study, we consider a 2D SD system positioned in the $x$-$y$ plane (taken as the
2D plane). The effective two band model Hamiltonian \cite{Pyatkovskiy16,Baik15,Mawrie19,Mawrie19R}
for a carrier (an electron or a hole) in reciprocal space can be written as
\begin{equation}\label{1}
 H(\mathbf{k})=\left(\begin{array}{cc}
  0 & \Delta+a k_{x}^2-i\hbar v_{\mathrm{F}}k_{y} \\
  \Delta+a k_{x}^2+i\hbar v_{\mathrm{F}}k_{y} & 0 \\
  \end{array}\right),
\end{equation}
where $\mathbf{k}=(k_x,k_y)$ is the carrier wave vector or momentum operator, the quadratic
dispersion is along the $x$ direction, $v_{\mathrm{F}}$ is the Fermi velocity,
$a=\hbar^2/(2m_x)$ with $m_x$ being the effective carrier mass along the $x$ direction,
$2\Delta$ is the band gap between the conduction and valence bands when
$\Delta>0$ and 2D SD would be a gapless system when $\Delta\leq0$ with $\Delta$ being the
band-gap parameter, $k_{\pm}=k_x\pm ik_y=ke^{\pm i\theta}$, and $\theta$ is the angle between
$\mathbf{k}$ and the $x$ axis.

The corresponding Schr\"{o}dinger equation for 2D SD system can be solved analytically.
The eigenvalues and eigenfunctions are given respectively by
\begin{equation}\label{energysp}
\varepsilon^{\lambda}_{\mathbf{k}}=\lambda\sqrt{[ak_{x}^2+\Delta]^2
+\hbar^2 v^2_{\mathrm{F}}k^2_{y}},
\end{equation}
where $\lambda=+/-$ refers to conduction/valence band, and
\begin{equation}
\psi^\lambda_{\mathbf{k}}(\mathbf{r})=\frac{1}{\sqrt{2}}\left(\begin{array}{c}
\lambda \\
 g_k \\
  \end{array}\right)
e^{i\mathbf{k}\cdot\mathbf{r}},
\end{equation}
with $$g_k={ak_{x}^2+\Delta+i\hbar v_{\mathrm{F}}k_{y}\over [(ak^2_{x}+\Delta)^2
+\hbar^2 v^2_{\mathrm{F}}k^2_{y}]^{1/2}}.$$

We now consider that an elliptically polarized CW radiation field \cite{Zhang18} is
applied normal to the 2D plane of the 2D SD system. Within the Coulomb gauge, the
vector potential of the light field is given by the Jones vector \cite{Pedrotti17}
with the left ($\nu=-$) and right ($\nu=+$) handed elliptically polarization, with an
ellipticity ratio $\eta$, is
\begin{align}
A^{0^\mathrm{o}}_{\nu}(t)=C
(\hat{\mathbf{x}}+i\nu\eta \hat{\mathbf{y}}),
\end{align}
where $C=F_0(1+\eta^2)^{-1/2}\sin(\omega t)/\omega$ with $F_0$ and $\omega$ being respectively
the electric field strength and the frequency of the radiation field. For the case of relatively
weak radiation, the carrier-photon interaction Hamiltonian can be written as
\begin{align}
H'_{\nu}(t)=eC\left(\begin{array}{cc}
  0 & q+ \nu p  \\
  q-\nu p & 0 \\
  \end{array}\right).
\end{align}
where $p=v_\mathrm{F}\eta$ and $q=\hbar k\cos\theta/m_x$. Thus, taking $H'_{\nu}$
as a perturbation and applying Fermi's golden rule, the first-order contribution
to the steady-state electronic transition rate induced by direct carrier-photon
interaction can be obtained as
\begin{align}
W^{\mp,0^\mathrm{o}}_{\lambda\lambda'}&(\mathbf{k},\mathbf{k'})=
\frac{\pi e^2F^2_0}{2\hbar\omega^2(1+\eta^2)}
\delta(\varepsilon^{\lambda'}_{\mathbf{k'}}
-\varepsilon^{\lambda}_{\mathbf{k}}
\mp\hbar\omega)\delta_{\mathbf{k},\mathbf{k'}}\nonumber\\
&\times[q^2\delta_{\lambda,\lambda'}+p^2\delta_{\lambda,-\lambda'}-\lambda\lambda'r(q^2-p^2)],
\end{align}
where $\mp$ sign in the Delta function refers to the absorption ($-$)
or emission ($+$) of a photon with energy $\hbar\omega$, and
$$r=\frac{\hbar^2 v^2_{\mathrm{F}}k^2\sin^2\theta}{[ak^2\cos^2\theta+\Delta]^2
+\hbar^2 v^2_{\mathrm{F}}k^2\sin^2\theta}.$$

After the rotation of the sample coordinates or of the light polarization by
$90^\mathrm{o}$, the vector potential of the radiation field now becomes
\begin{align}
A^{90^\mathrm{o}}_{\nu}(t)=C(\hat{\mathbf{y}}+i\nu\eta \hat{\mathbf{x}}),
\end{align}
The carrier-photon interaction Hamiltonian is then given by
\begin{align}
H'_{\nu}(t)=ieC\left(\begin{array}{cc}
  0 & \nu \eta q -v_{\mathrm{F}} \\
  \nu \eta q+v_{\mathrm{F}} & 0 \\
  \end{array}\right),
\end{align}
and the corresponding electronic transition rate is obtained as
\begin{align}
W^{\mp,90^\mathrm{o}}_{\lambda\lambda'}&(\mathbf{k},\mathbf{k'})=
\frac{\pi e^2F^2_0}{2\hbar\omega^2(1+\eta^2)}
\delta(\varepsilon^{\lambda'}_{\mathbf{k'}}
-\varepsilon^{\lambda}_{\mathbf{k}}
\mp\hbar\omega)\delta_{\mathbf{k},\mathbf{k'}}\nonumber\\
&\times[\eta^2q^2\delta_{\lambda,\lambda'}+v^2_{\mathrm{F}}\delta_{\lambda,-\lambda'}
-\lambda\lambda'r  
(\eta^2q^2-v^2_{\mathrm{F}})].
\end{align}

In this paper, we use the Boltzmann equation approach to study the
response of the carriers in a 2D SD system to the applied radiation
field. For nondegenerate statistics, the semiclassical Boltzmann
equation takes the form \cite{Xu10}
\begin{align}
\frac{\partial f_{\bf k}^\lambda}{\partial t}=&g_s \sum_{\lambda',\mathbf{k'}}
[W^-_{\lambda'\lambda}(\mathbf{k'},\mathbf{k})
f^{\lambda'}_\mathbf{k'}(1-f^\lambda_\mathbf{k})\nonumber\\
&-W^-_{\lambda\lambda'}(\mathbf{k},\mathbf{k'})
f^{\lambda}_\mathbf{k}(1-f^{\lambda'}_\mathbf{k'})],
\end{align}
where $g_s=2$ counts for the spin degeneracy, $f^{\lambda}_\mathbf{k}\simeq f_{\lambda}(\varepsilon^{\lambda}_{\mathbf{k}})
=\{\exp[(\varepsilon^\lambda_{\mathbf{k}}-\mu_\lambda)/(k_BT)]
+1\}^{-1}$
is the statistical energy distribution for the carriers such as the
Fermi-Dirac function, and $\mu_\lambda$ is the
chemical potential (or Fermi energy $\varepsilon^\lambda_{\mathrm{F}}$
at zero temperature) for electrons or holes in conduction or valence
bands. Taking the first moment, the energy-balance equation can be derived by multiplying $g_s \sum_{\mathbf{k},\lambda}E_\lambda(\mathbf{k})$ to both sides of the
Boltzmann equation \cite{Xu10}. From the energy-balance equation, we obtain the
energy transfer rate:
\begin{equation}
P_\eta=4\hbar\omega \sum_{\lambda',\lambda}\sum_{\mathbf{k'},\mathbf{k}}
W^-_{\lambda\lambda'}(\mathbf{k},\mathbf{k'})f^{\lambda}_\mathbf{k}
(1-f^{\lambda'}_\mathbf{k'}),
\end{equation}
where $P_\eta=\partial [g_s\sum_{\mathbf{k},\lambda}E_\lambda(\mathbf{k})f^{\lambda}_\mathbf{k}] /\partial t$.
The optical conductivity can be obtained by \cite{Xu10}: $\sigma^\eta(\omega)=2P_\eta/F_0^2$
which have intra- and interband electronic transition channels
\begin{equation}
\sigma^\eta(\omega)=\sum_{\lambda',\lambda}\sigma^\eta_{\lambda\lambda'}(\omega).
\end{equation}

We now consider an $n$-type 2D SD system irradiated by an elliptically polarized CW
radiation field. When photon energy $\hbar\omega$ is larger than the band gap
between the conduction and valence bands, the photo-excited carriers are induced in the
conduction and valence bands and a quasi-equilibrium state is established in
the system. In this case, the electron density is $n_e=n_0+\Delta n$ with
$n_0$ being the dark electron density and $\Delta n$ the photo-induced electron density.
Due to charge conservation, the hole density in the system is $n_h=\Delta n$.
Furthermore, the channels of the optical absorption can be induced by intraband electronic
transitions within the conduction and valence bands via the mechanism of
free-carrier absorption and by interband transitions from valence band to conduction band.
The interband electronic transition from conduction band to valence band via optical
absorption is physically impossible. Taking the real part of the vector potential induced
by polarized radiation field along the $xx$ direction, the optical conductivity via different
transition channels can be obtained as
\begin{align}\label{cond01}
\sigma^{\eta, 0^\mathrm{o}}_{++}(\omega)=&\frac{e^2 \delta(\hbar\omega)}{2\pi\omega(1+\eta^2)}\int_0^{2\pi}d\theta\int_0^\infty
dkk[q^2-r(q^2-p^2)]\nonumber\\
&\times f_{+}(\varepsilon^{+}_{\mathbf{k}})
[1-f_{+}(\varepsilon^{+}_{\mathbf{k}})],
\end{align}
\begin{align}\label{cond02}
\sigma^{\eta, 0^\mathrm{o}}_{--}(\omega)=&\frac{e^2 \delta(\hbar\omega)}{2\pi\omega(1+\eta^2)}\int_0^{2\pi}d\theta\int_0^\infty
dkk[q^2-r(q^2-p^2)]\nonumber\\
&\times f_{-}(\varepsilon^{-}_{\mathbf{k}})
[1-f_{-}(\varepsilon^{-}_{\mathbf{k}})],
\end{align}
and
\begin{align}\label{cond03}
\sigma^{\eta, 0^\mathrm{o}}_{-+}(\omega)=&\frac{e^2}{2\pi\omega(1+\eta^2)}\int_0^{2\pi}d\theta\int_0^\infty
dkk[p^2+r(q^2-p^2)]\nonumber\\
&\times f_{-}(\varepsilon^{-}_{\mathbf{k}})
[1-f_{+}(\varepsilon^{+}_{\mathbf{k}})]\delta(\varepsilon^{+}_{\mathbf{k}}
-\varepsilon^{-}_{\mathbf{k}}
-\hbar\omega),
\end{align}
respectively.

Here, the $\delta$ function will be replaced by a Lorentzian distribution
under the energy relaxation approximation for intraband transitions: $\delta(\varepsilon)\rightarrow(\varepsilon_\tau/\pi)/(\varepsilon^2+\varepsilon^2_\tau)$,
where $\varepsilon_\tau=\hbar/\tau$ is the width of the broadened energy state
with $\tau$ being the energy relaxation time.

Taking the real part of the vector potential induced by the polarized radiation
field along the $yy$ direction, the optical conductivity via the different transition
channels are given as
\begin{align}\label{cond901}
\sigma^{\eta,90^\mathrm{o}}_{++}(\omega)=&\frac
{e^2 \delta(\hbar\omega)}{2\pi\omega(1+\eta^2)}\int_0^{2\pi}d\theta\int_0^\infty
dkkf_{+}(\varepsilon^{+}_{\mathbf{k}})\nonumber\\
&\times
[1-f_{+}(\varepsilon^{+}_{\mathbf{k}})][\eta^2q^2-r(\eta^2q^2-v^2_{\mathrm{F}})],
\end{align}
\begin{align}\label{cond902}
\sigma^{\eta,90^\mathrm{o}}_{--}(\omega)=&\frac{ e^2 \delta(\hbar\omega)}{2\pi\omega(1+\eta^2)}\int_0^{2\pi}d\theta\int_0^\infty
dkkf_{-}(\varepsilon^{-}_{\mathbf{k}})\nonumber\\
&\times
[1-f_{-}(\varepsilon^{-}_{\mathbf{k}})][\eta^2q^2-r(\eta^2q^2-v^2_{\mathrm{F}})],
\end{align}
and
\begin{align}\label{cond903}
\sigma^{\eta,90^\mathrm{o}}_{-+}(\omega)=&\frac{ e^2}{2\pi\omega(1+\eta^2)}
\int_0^{2\pi}d\theta\int_0^\infty
[v^2_{\mathrm{F}}+r(\eta^2q^2-v^2_{\mathrm{F}})]\nonumber\\
&\times f_{-}(\varepsilon^{-}_{\mathbf{k}})
[1-f_{+}(\varepsilon^{+}_{\mathbf{k}})]\delta(\varepsilon^{+}_{\mathbf{k}}
-\varepsilon^{-}_{\mathbf{k}}
-\hbar\omega)kdk.
\end{align}
From Eqs. \eqref{cond01}--\eqref{cond903},
we can obtain the following relations with
\begin{equation}
\sigma^{\eta,0^\mathrm{o}}_{\lambda\lambda'}(\omega)
=\frac{\sigma^{\lambda\lambda'}_{xx}(\omega)
+\eta^2\sigma^{\lambda\lambda'}_{yy}(\omega)}{1+\eta^2},
\end{equation}
and
\begin{equation}
\sigma^{\eta,{90}^\mathrm{o}}_{\lambda\lambda'}(\omega)
=\frac{\eta^2\sigma^{\lambda\lambda'}_{xx}(\omega)
+\sigma^{\lambda\lambda'}_{yy}(\omega)}{1+\eta^2},
\end{equation}
where $\sigma^{\lambda\lambda'}_{xx}(\omega)$ and
$\sigma^{\lambda\lambda'}_{yy}(\omega)$ are the longitudinal
optical conductivities in $xx$ and $yy$ directions for different
electronic transition channels.
The interband part of the longitudinal optical conductivity
$\sigma^{-+}_{xx}(\omega)$ and $\sigma^{-+}_{yy}(\omega)$ can
be written respectively as
\begin{align}\label{condxx}
\sigma_{xx}^{-+}(\omega)=&\frac{e^2}{\pi\hbar\omega^2}
\int_0^{\pi/2}d\theta M(\theta)f_{-}(-\frac{\hbar\omega}{2})[1-f_{+}(\frac{\hbar\omega}{2})],
\end{align}
where
\begin{align}
M(\theta)=\bigg\{\begin{array}{cc}
I(k^2_{\theta0}),&\cos\theta=0\\
I(k^2_{\theta+})+I(k^2_{\theta-}), &\cos\theta\neq0\\
  \end{array}
\end{align}
with
\begin{align}
I(k_{\theta\varsigma}^2)=\frac{2a^2v^2_\mathrm{F}k_{\theta\varsigma}^4\sin^2(2\theta)\Theta(k^2_{\theta\varsigma})}
{|2a\cos^{2}\theta(ak_{\theta\varsigma}^2\cos^2\theta+\Delta)+\hbar^2v^2_\mathrm{F}\sin^2\theta|
}.\nonumber
\end{align}
The step functions is defined as
\begin{equation}
\Theta(x)=\bigg\{
\begin{array}{c}
1 ,x>0\\
0 ,x\leq0\\
\end{array},
\end{equation}
and $k^2_{\theta\varsigma}$ should be real and positive, which writes
\begin{align}
k^2_{\theta\varsigma}=\bigg\{\begin{array}{cc}
 (\hbar^2\omega^2-4\Delta^2)/(4\hbar^2v_\mathrm{F}^2),&\cos\theta=0,\varsigma=0\\
(\varsigma \sqrt{V}-U)/(2a^2\cos^4\theta), &\cos\theta\neq0,\varsigma=\pm\\
  \end{array}
\end{align}
where
$U=2\Delta a\cos^2\theta+\hbar^2v_\mathrm{F}^2\sin^2\theta$ and
$V=U^2+a^2\cos^4\theta(\hbar^2\omega^2-4\Delta^2)$.
Moreover, The interband part of the longitudinal optical conductivity
in $yy$ direction $\sigma^{-+}_{yy}(\omega)$ is given by
\begin{align}\label{condyy}
\sigma_{yy}^{-+}(\omega)=\frac{e^2}{\pi\hbar\omega^2}
\int_0^{\pi/2}d\theta N(\theta)f_{-}(-\frac{\hbar\omega}{2})
[1-f_{+}(\frac{\hbar\omega}{2})],
\end{align}
where
\begin{align}
N(\theta)=\bigg\{\begin{array}{cc}
J(k^2_{\theta0}),&\cos\theta=0\\
J(k^2_{\theta+})+J(k^2_{\theta-}), &\cos\theta\neq0\\
  \end{array}
\end{align}
with
\begin{align}
J(k^2_{\theta\varsigma})=\frac{2v^2_\mathrm{F}(ak^2_{\theta\varsigma}
\cos^2\theta+\Delta)^2\Theta(k^2_{\theta\varsigma})}
{|2a\cos^{2}\theta(ak_{\theta\varsigma}^2\cos^2\theta+\Delta)+\hbar^2v^2_\mathrm{F}\sin^2\theta|
}.\nonumber
\end{align}

\section{Results and discussions}
\label{sec:results}

\begin{figure}[b]
\centering
\includegraphics[width=8.2cm]{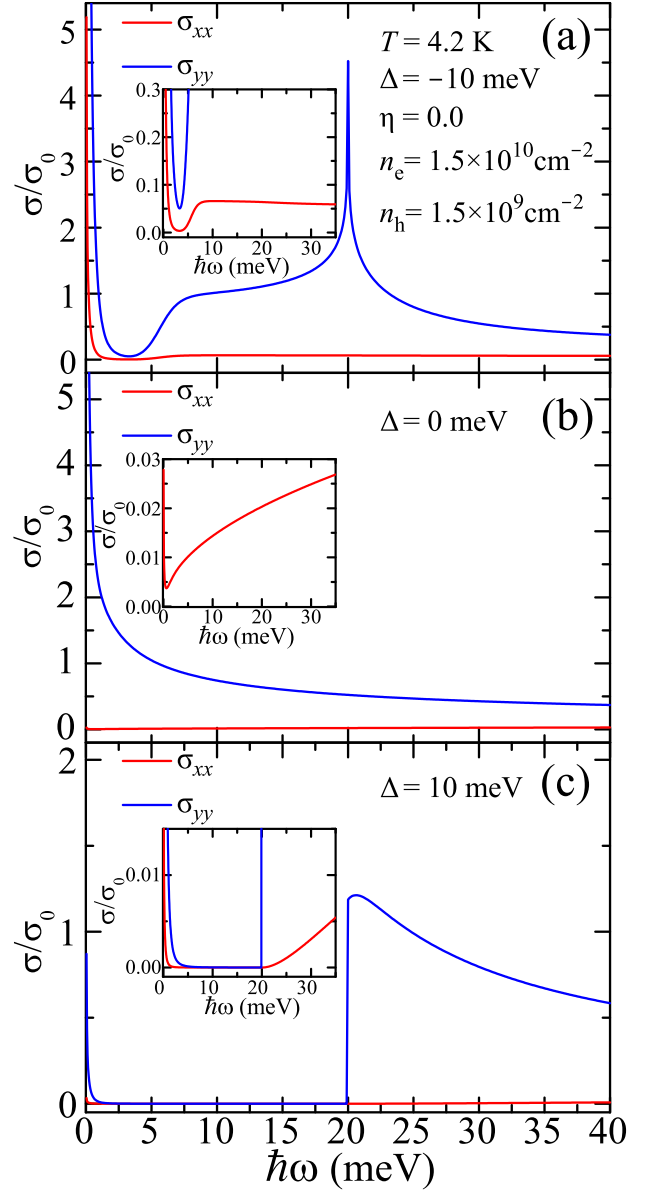}
\caption{Longitudinal optical conductivities
$\sigma_{xx}(\omega)$ and $\sigma_{yy}(\omega)$ as a
function of radiation photon energy $\hbar\omega$ at the fixed
band-gap parameter $\Delta=-10$ meV (a), $\Delta=0$ meV (b), and
$\Delta=10$ meV (c), with temperatures $T=4.2$ K, ellipticity ratio $\eta=0$,
and carrier densities $n_e=1.5\times10^{10}$ cm$^{-2} $ and $n_h=1.5\times10^9$ cm$^{-2}$.
The insets in (a--c) show zoom in of $\sigma_{xx}(\omega)$ and $\sigma_{yy}(\omega)$.
Here $\sigma_0=e^2/\hbar$.}
\label{fig1}
\end{figure}

For our numerical calculations, we take the following typical material
parameters obtained from literatures: $m_x=5.08m_e$, where $m_e$ is the
rest electron mass, and the Fermi velocity $v_\mathrm{F}=9.875\times10^4$
m/s \cite{Saha16,Mawrie19,Mawrie19R}.
However, the material parameters would differ within different materials
such as $m_x=13.6m_e$, $v_\mathrm{F}=1.5\times10^5$ m/s in (TiO$_2$)$_5$/(VO$_2$)$_3$
and $m_x=3.1m_e$, $v_\mathrm{F}=1.14\times10^5$ m/s in $\alpha$-(BEDT-TTF)$_2$I$_3$ \cite{Saha17}.
We use the electronic relaxation time $\tau=1.0$ ps for calculations of the
optical conductivity induced by intraband electronic transitions. The chemical
potentials (or Fermi energies) in the conduction and valence bands in a 2D SD system
can be determined respectively by using the condition of carrier number conservation with
the given electron and hole densities $n_\lambda$ ($\lambda=+$ for conduction band
and $\lambda=-$ for valence band) through
\begin{equation}\label{Cdensi}
n_{\lambda}=\frac{g_s}{(2\pi)^2}\int_{0}^{2\pi}d\phi\int_{0}^{\infty}  dkk
[\delta_{\lambda,-1}+\lambda f_\lambda(\varepsilon^{\lambda}_{\mathbf{k}})],
\end{equation}
where $f_{\lambda}(\varepsilon^{\lambda}_{\mathbf{k}})$ is the Fermi-Dirac
function for electrons in conduction band or holes in valence band.

It should be noted that the electronic relaxation time $\tau$
depends on the carrier densities and temperature via electron-electron
and electron-phonon interactions. We take a phenomenological parameter for showing the
features of the intraband optical conductivity. The relaxation time $\tau$ could
be obtained from experiments via, e.g., transport measurement \cite{Bandurin16} or ultra fast
optical spectroscopy \cite{Sun08}. By the way, in the framework of Fermi's golden rule, the optical
conductivities obtained in this study corresponds to the optical absorption in a relative
weak elliptically polarized radiation because a strong nonlinear optical response in a 2D
SD system would be occurred in the strong terahertz radiation \cite{Dai19,Samal21}.

\begin{figure}[t]
\centering
\includegraphics[width=8.6cm]{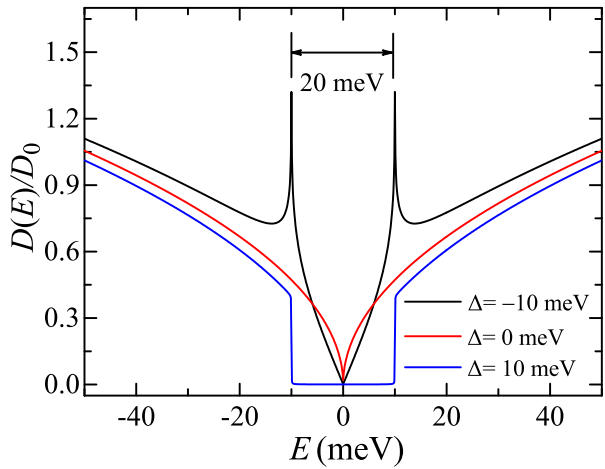}
\caption{The density-of-states (DoS) per unit area for a 2D SD system with different band-gap parameters
$\Delta=-10$ meV (black curve), $\Delta=0$ meV (red curve), and $\Delta=10$ meV (blue curve),
respectively. Here $D_0=10^{19}$/meV.}\label{fig2}
\end{figure}

\begin{figure}[t]
\centering
\includegraphics[width=8.6cm]{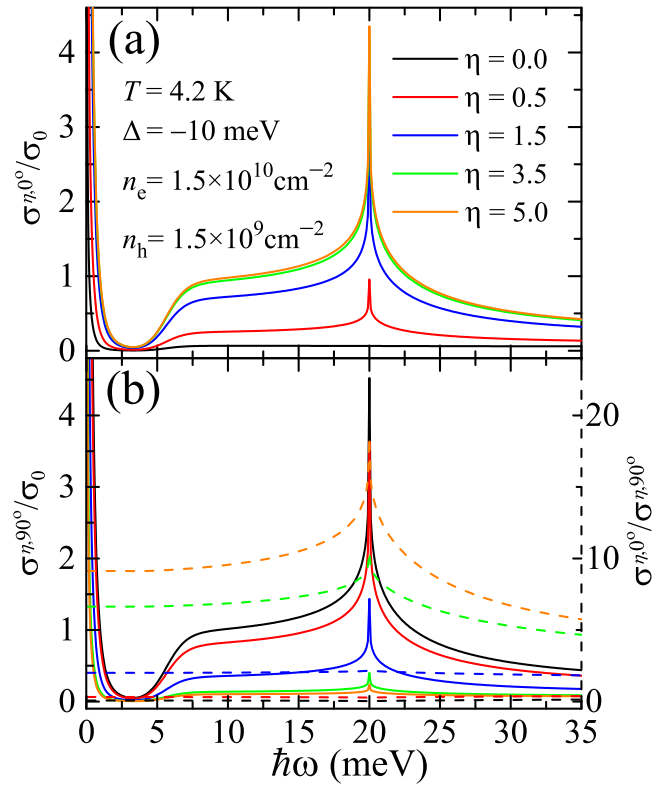}
\caption{Optical conductivities (a) $\sigma^{\eta,0^\mathrm{o}}(\omega)$ and (b)
$\sigma^{\eta,90^\mathrm{o}}(\omega)$ as a function of photon energy
$\hbar\omega$ at the fixed band gap parameter $\Delta=-10$ meV, temperature
$T=4.2$ K, carrier densities $n_e=1.5\times10^{10}$ cm$^{-2}$, and
$n_h=1.5\times10^9$ cm$^{-2}$ for different ellipticity ratio $\eta=0.0$ (black curve),
$0.5$ (red curve), $1.5$ (blue curve), $3.5$ (green curve), and $5.0$ (orange curve),
respectively. The dashed curves in (b) are the corresponding ratios of $\sigma^{\eta,0^\mathrm{o}}(\omega)/\sigma^{\eta,90^\mathrm{o}}(\omega)$
for the different ellipticity ratios.}\label{fig3}
\end{figure}

\begin{figure}[t]
\centering
\includegraphics[width=8.6cm]{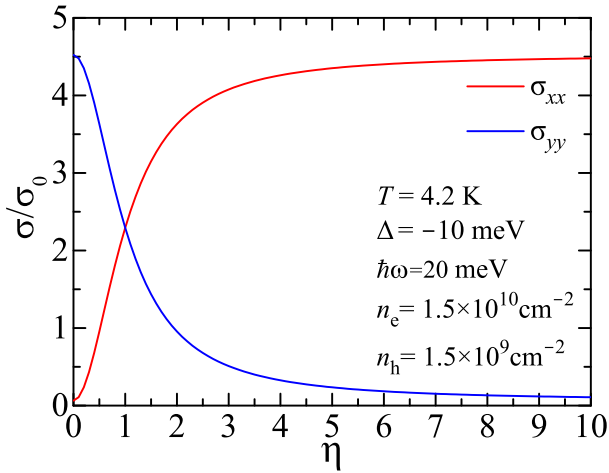}
\caption{The hight of the absorption peak in Fig. \ref{fig3}
at a fixed photon energy 20 meV as a function of ellipticity ratio $\eta$ with
band-gap parameter $\Delta=-10$ meV, temperature $T=4.2$ K, carrier densities
$n_e=1.5\times10^{10}$ cm$^{-2}$, and $n_h=1.5\times10^9$ cm$^{-2}$.}\label{fig4}
\end{figure}

\begin{figure}[t]
\centering
\includegraphics[width=8.6cm]{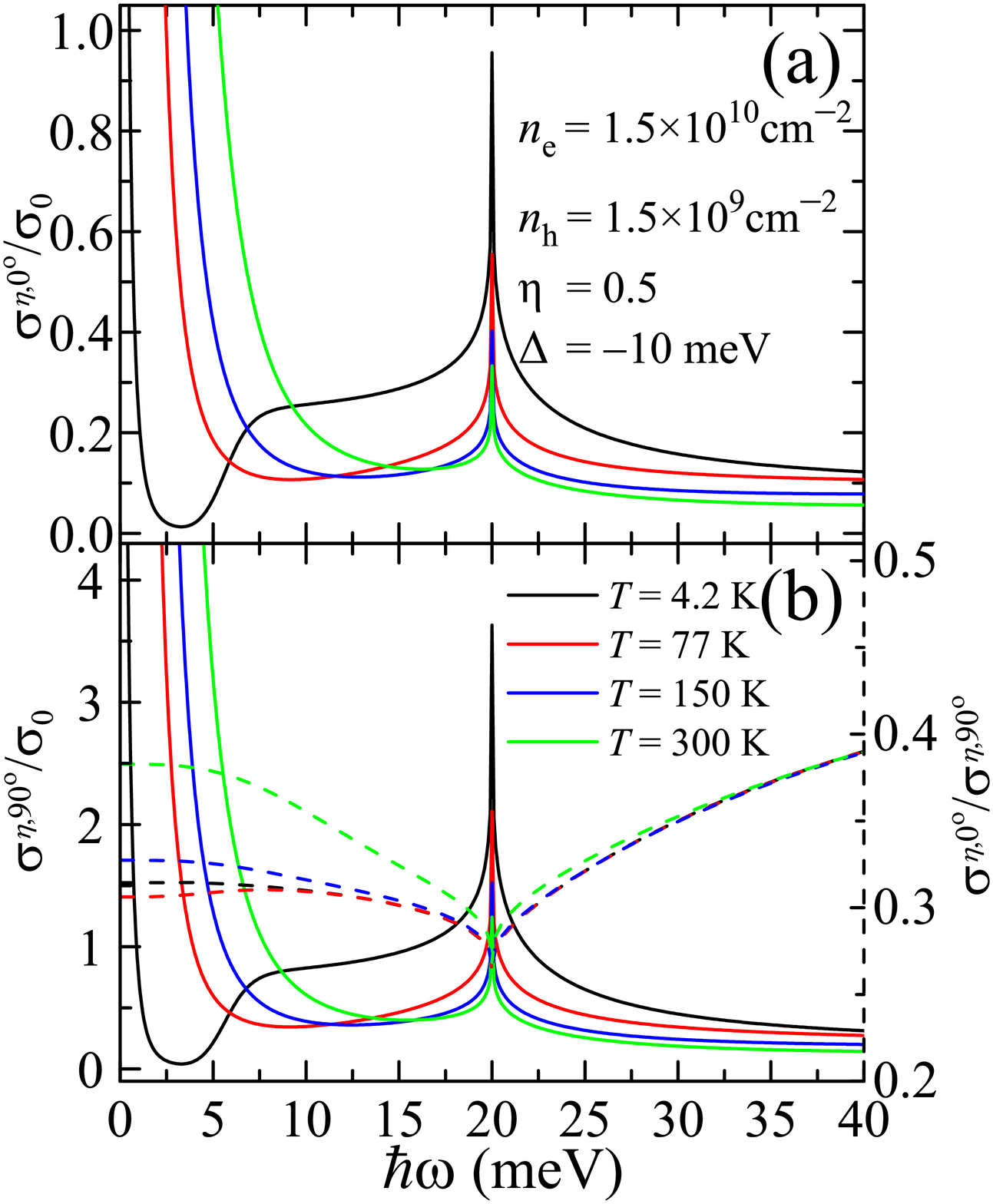}
\caption{Optical conductivities (a) $\sigma^{\eta,0^\mathrm{o}}(\omega)$ and (b)
$\sigma^{\eta,90^\mathrm{o}}(\omega)$ as a function of photon energy
$\hbar\omega$ at the fixed band-gap parameter $\Delta=-10$ meV, ellipticity
ratio $\eta=0.5$, carrier densities $n_e=1.5\times10^{10}$ cm$^{-2}$, and
$n_h=1.5\times10^9$ cm$^{-2}$ for different temperatures $T=4.2$ K (black curve),
$77$ K (red curve), $150$ K (blue curve), and $300$ K (green curve),
respectively. The dashed curves in (b) are the corresponding ratios of $\sigma^{\eta,0^\mathrm{o}}(\omega)/\sigma^{\eta,90^\mathrm{o}}(\omega)$
for different temperatures.}\label{fig5}
\end{figure}

In Fig. \ref{fig1}, we plot the optical conductivity
$\sigma^{\eta, 0^\mathrm{o}}(\omega)$ and $\sigma^{\eta, 90^\mathrm{o}}(\omega)$
as a function of photon energy $\hbar\omega$ at fixed ellipticity ratio, carriers
density, temperature, and band-gap parameter. The total optical conductivity is
due to both the inter- and the intraband transition channels. For $\eta=0$, we obtain
that the longitudinal optical conductivities along different directions are $\sigma^{0,0^\circ}(\omega)=\sigma_{xx}(\omega)$ and
$\sigma^{0,{90}^\circ}(\omega)=\sigma_{yy}(\omega)$. The optical absorption is
anisotropic for both interband transitions and intraband transitions.
From Eqs. \eqref{cond03} and \eqref{cond903}, we learn that the optical conductivities
are independent upon the sign index $\nu=\pm1$ for left- or right-handed elliptically
polarized light, which means that the transverse or Hall optical conductivity in a 2D
SD system is zero. In Figs. \ref{fig1}(a-c), the longitudinal optical conductivity
$\sigma_{yy}(\omega)$ is larger than $\sigma_{xx}(\omega)$ in the whole spectrum regime.
At high radiation frequencies, the optical absorption in the $y$ direction is significant
stronger than that in the $x$ direction, which is in line with the results obtained
previously \cite{Mawrie19,Carbotte19,Carbotte19s}. In the low-frequency regime, both
$\sigma_{xx}(\omega)$ and $\sigma_{yy}(\omega)$ decrease monotonously with increasing
$\omega$, a typical feature of the Drude-like optical conductivity for free carriers \cite{Xiao16}.
It should be noted that in Refs. \cite{Carbotte19,Carbotte19s}, Carbotte $et\ al$. had
also calculated the longitudinal optical conductivity of 2D semi-Dirac system in $xx$ and $yy$
directions with/without a gap within a Kubo formalism. Carbotte $et\ al$. provided separate
analytic formulas of optical conductivity for intraband and interband transitions
for certain limitation cases and considered the transport properties such as dc conductivity,
thermal conductivity, and the Lorenz number. In Fig. \ref{fig1}(a), we note that an absorption
peak can be observed at a photon energy $\hbar\omega=20$ meV. In Fig. \ref{fig1}(b), the
interband optical conductivities $\sigma_{xx}(\omega)$ and $\sigma_{yy}(\omega)$ at low-temperature behavior
as $\sqrt{\omega}$ and $1/\sqrt{\omega}$, respectively, as obtained in Ref. \cite{Carbotte19}.
In Fig. \ref{fig1}(c), we can see the cutoff in 20 meV is due to the prohibition of interband transitions
below the band gap. In general, the results of longitudinal optical conductivities
shown in Fig. \ref{fig1} are in line with the results obtained in Refs. \cite{Carbotte19,Carbotte19s,Mawrie19}.
The van Hove singularity of $\sigma_{yy}(\omega)$ in Fig. \ref{fig1}(a) can be understood from the electronic
density-of-states (DoS). In Fig. \ref{fig2}, we show the DoS for a 2D
SD system with different band-gap parameters. The DoS per unit area is determined by the imaginary part
of the Green's function via
\begin{equation}
D(E)=\frac{g_s}{\pi}\sum_{\lambda=\pm,\mathbf{k}}\mathrm{Im}\
G_{\lambda}(E)=g_s\sum_{\lambda=\pm,\mathbf{k}}\delta(E-\epsilon^{\lambda}_\mathbf{k}),
\end{equation}
where $G_{\lambda}(E)=[E-\epsilon^{\lambda}_k+ i\delta]^{-1}$ is the
retarded Green's functions for a carrier in the $\lambda=\pm$ band and $E$ is the electron
energy. We take a broadened width of $\varepsilon_\tau=1.0$ meV with replacing the
delta function with energy relation approximation.
From Fig. \ref{fig2}, we see that the DoS for a 2D SD system differs significantly
from that for a semiconductor-based 2D electron gas, which is a step-function, and from
that for graphene, which is linear. The DoS for the 2D SD system depends strongly on
the band-gap parameter $\Delta$. For a band-gap parameter of $\Delta=-10$ meV, we see that
the energy spacing between the two peaks in DoS is 20 meV, which corresponds to an electronic
transition energy of 20 meV and, thus, results in a strong absorption peak in optical
conductivity as shown in Fig. \ref{fig1}(a). For the case of $\Delta \ge 0$ the DoS for a 2D
SD system increases nonlinearly with $E$ starting from the band edges. As a result, the
interesting features of the DoS for a 2D SD system are the main reasons why the corresponding
optoelectronic properties are different from semiconductor-based 2D electron gas systems
and from graphene.

\begin{figure}[t]
\centering
\includegraphics[width=8.6cm]{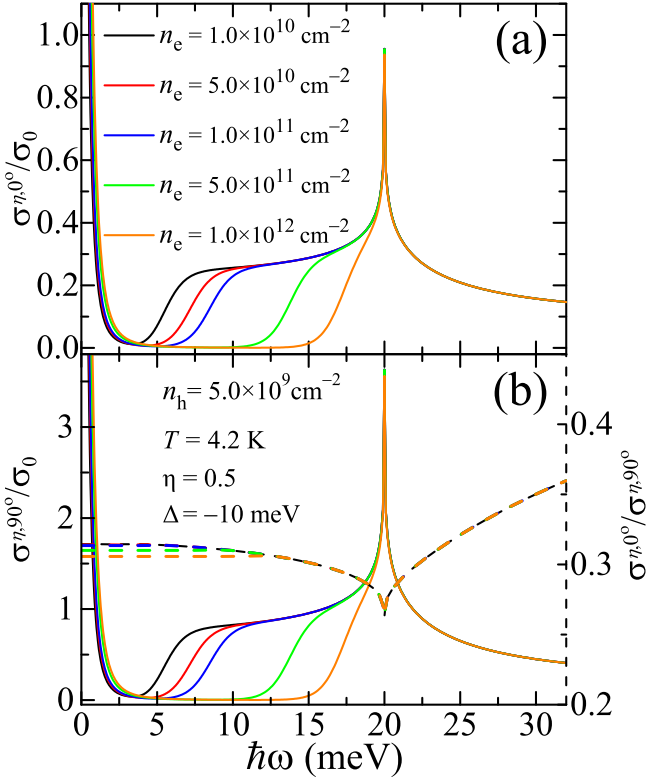}
\caption{Optical conductivities (a) $\sigma^{\eta,0^\mathrm{o}}(\omega)$ and (b)
$\sigma^{\eta,90^\mathrm{o}}(\omega)$ as a function of photon energy
$\hbar\omega$ at the fixed band-gap parameter $\Delta=-10$ meV, ellipticity ratio
$\eta=0.5$, temperature $T=4.2$ K, hole density $n_h=5.0\times10^9$ cm$^{-2}$
with different electron densities $n_e=1.0\times10^{10}$ cm$^{-2}$ (black curve),
$5.0\times10^{10}$ cm$^{-2}$ (red curve), $1.0\times10^{11}$ cm$^{-2}$ (blue curve),
$5.0\times10^{11}$ cm$^{-2}$ (green curve), and $1.0\times10^{12}$ cm$^{-2}$
(orange curve), respectively. The dashed curves
in (b) are the corresponding ratios of $\sigma^{\eta,0^\mathrm{o}}(\omega)
/\sigma^{\eta,90^\mathrm{o}}(\omega)$ for different electron densities as indicated.}\label{fig6}
\end{figure}

In Fig. \ref{fig3}, the optical conductivities $\sigma^{\eta,0^\mathrm{o}}(\omega)$ in
(a) and $\sigma^{\eta,90^\mathrm{o}}(\omega)$ in (b) are shown as a function of radiation
photon energy at a fixed band-gap parameter, temperature, carrier densities for different
values of the ellipticity ratios $\eta$. With increasing ellipticity ratio, the values
of the optical conductivity in longitudinal direction
$\sigma^{\eta,0^\mathrm{o}}(\omega)=\sigma_{xx}^\eta (\omega)$ increases in the whole spectral
regime and the optical conductivity $\sigma^{\eta,90^\mathrm{o}}(\omega)=\sigma_{yy}^\eta (\omega)$
in the vertical direction decreases. From Fig. \ref{fig3}(b), we can also see that the ratio
of the optical conductivity between longitudinal and vertical directions, $\sigma^{\eta,0^\mathrm{o}}(\omega)/\sigma^{\eta,90^\mathrm{o}}(\omega)$,
increases with increasing $\eta$. In Fig. \ref{fig4}, we plot
the peak hight of the optical conductivity in Fig. \ref{fig3} at a
photon energy 20 meV as a function of
ellipticity ratio. We can see that the peak hight of $\sigma^{\eta,0^\mathrm{o}}(\omega)/
\sigma^{\eta,90^\mathrm{o}}(\omega)$ increases/decreases with increasing the ellipticity ratio.
The absorption peak hight in van Hove singularity can be effectively tuned
by the ellipticity ratio. In the presence of elliptically polarized light
with different ellipticity ratios, the 2D SD system shows a strong anisotropy on the
optical absorption. Hence, the ellipticity ratio of light
radiation can effectively tune the optical and optoelectronic properties of the 2D SD system
in the infrared and THz regime.

In Fig. \ref{fig5}, we show the effect of temperature on the optical conductivity
spectrum of the 2D SD system at the fixed band-gap parameter, ellipticity ratio,
and carrier densities. In the long-wavelength regime where the free-carrier absorption
contributes mainly to the intra-band transitions, the optical absorption increases with
increasing temperature. In the intermediate frequency regime, the absorption near
the absorption peak is stronger at low temperature. For photon energies $\hbar\omega>35$
meV, the effect of temperature on the optical conductivity or absorption is weak.
However, the ratio of the optical conductivities in different directions $\sigma^{\eta,0^\mathrm{o}}(\omega)/\sigma^{\eta,90^\mathrm{o}}(\omega)$
becomes larger at higher temperatures, which implies that anisotropic optical absorption
can already be observed at room temperature.

\begin{figure}[t]
\centering
\includegraphics[width=8.6cm]{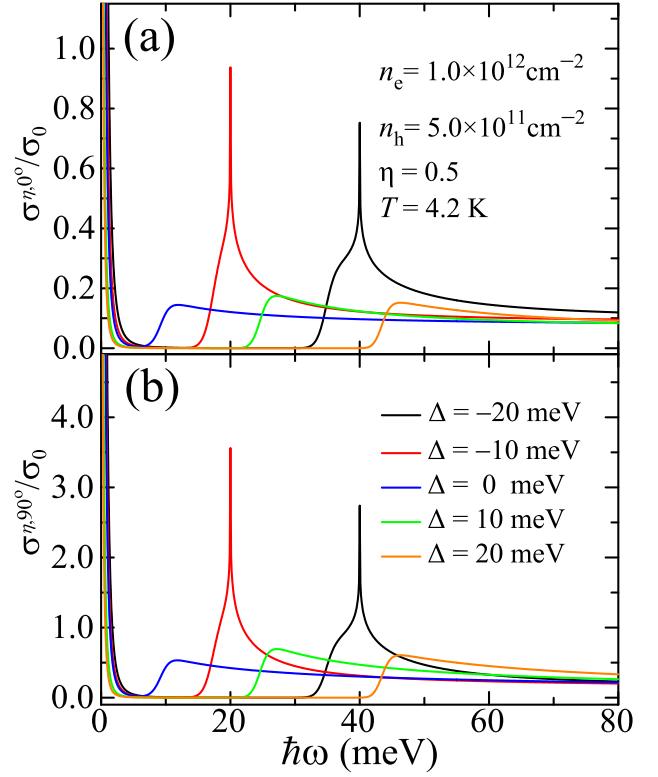}
\caption{Optical conductivities (a) $\sigma^{\eta,0^\mathrm{o}}(\omega)$ and (b)
$\sigma^{\eta,90^\mathrm{o}}(\omega)$ as a function of radiation photon energy
$\hbar\omega$ at the fixed ellipticity ratio $\eta=0.5$, temperature $T=4.2$ K,
carriers densities $n_e=1.0\times10^{12}$ cm$^{-2}$ and $n_h=5.0\times10^{11}$ cm$^{-2}$
with different band-gap parameters $\Delta=-20$ meV (black curve),
$-10$ meV (red curve), 0 meV (blue curve), 10 meV (green curve),
 and  20 meV (orange curve), respectively.}\label{fig7}
\end{figure}

In Fig. \ref{fig6}, we plot the optical conductivities $\sigma^{\eta,0^\mathrm{o}}(\omega)$ and
$\sigma^{\eta,90^\mathrm{o}}(\omega)$ as a function of photon energy at the fixed
band-gap parameter, temperatures, ellipticity ratio, hole density for different
electron densities. With increasing the electron density, the absorption edges in
the optical conductivities or absorption spectra show a blue shift and
the absorption at high frequency is not affected by the variation of the electron
density. This phenomenon is a result of the well known Pauli-blocking
effect \cite{Grigorenko12}, which indicates that electronic transitions can only occur
from occupied states to empty states at low temperatures. With increasing
electron density, the Fermi energy in conduction band increases. This can
increase the energy separation between the occupied valence band states and the empty
conduction band states and, thus, result in the blue shift in the optical absorption edge.
Similar to conventional electronic devices, the carrier density and chemical potential
in a 2D SD system can also be tuned, e.g., by applying a gate voltage. As can be see,
the absorption regime becomes narrow with increasing electron density. Thus, a wider
absorption spectrum can be observed in lower-density samples.

In Fig. \ref{fig7}, we show the optical conductivities
$\sigma^{\eta,0^\mathrm{o}}(\omega)$ and
$\sigma^{\eta,90^\mathrm{o}}(\omega)$ as a function of photon energy at the
fixed temperature, ellipticity ratio, carrier densities for different band-gap
parameters. With changing the band gap parameter, the
strength and the position of the optical absorption in infrared and THz regime can be
effectively tuned in the presence of an elliptically polarized light field.
From Eq. \eqref{energysp}, we know that the 2D SD system has a band-gap when
$\Delta>0$. At the fixed carrier concentrations, the absorption edge in Fig. \ref{fig7}
exhibits a blue shift for larger band-gap parameter. For a negative band-gap parameter
$\Delta<0$, the 2D SD system is gapless and presents unique
two nodal points in the energy spectrum \cite{Mawrie19,Pyatkovskiy16}. The absorption edge
is also blue shifted with decreasing $\Delta$. It has been shown that there exists
a widely tunable band gap in few-layer black phosphorus doped with potassium
using an $in\ situ$ surface doping technique \cite{Kim15} or by the induction of strain \cite{Kim21}.
Using such band-gap engineering approaches \cite{Kim15,Chaves20}, the anisotropic
optical absorption can be effectively tuned in the infrared and THz bandwidths.

\begin{figure}[t]
\centering
\includegraphics[width=8.6cm]{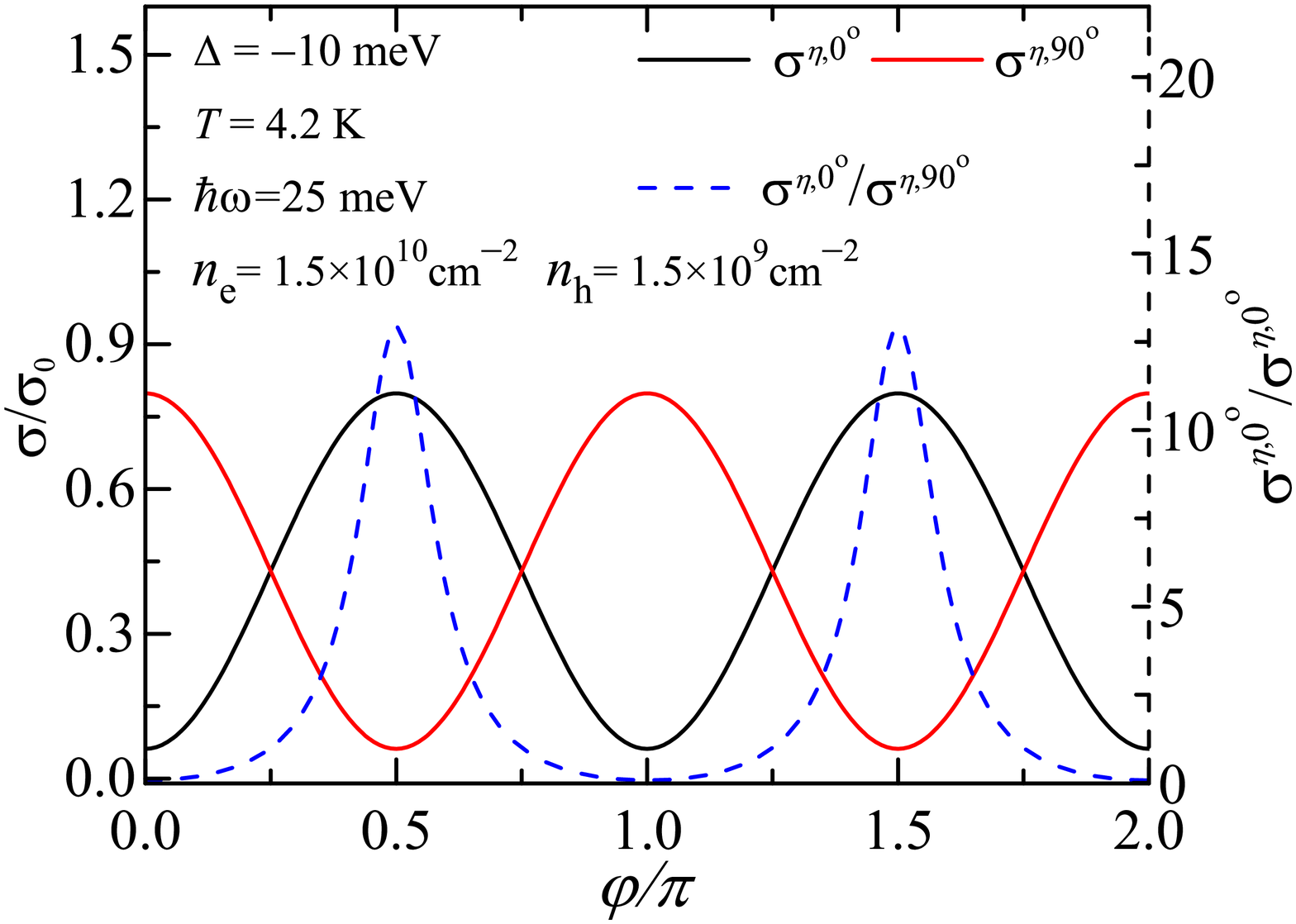}
\caption{Optical conductivities $\sigma^{\eta,0^\mathrm{o}}(\omega)$,
$\sigma^{\eta,90^\mathrm{o}}(\omega)$ and their ratios as a function
of ellipticity angle $\varphi$ ($\eta=\tan\varphi$) at fixed photon energy $\hbar\omega=25$ meV,
temperature $T=4.2$ K, band-gap parameter $\Delta=-10$ meV, carriers
densities $n_e=1.5\times10^{10}$ cm $^{-2}$ and $n_h=1.0\times10^9$ cm$^{-2}$.
The dashed curve is the ratio of $\sigma^{\eta,0^\mathrm{o}}
/\sigma^{\eta,90^\mathrm{o}}$.}\label{fig8}
\end{figure}

In Fig. \ref{fig8}, we plot the optical conductivities $\sigma^{\eta,0^\mathrm{o}}$ and
$\sigma^{\eta,90^\mathrm{o}}$ along different directions and their ratio as a function
of the ellipticity angle $\varphi$ at the fixed radiation photon energy $\hbar\omega=25$ meV,
temperature, band-gap parameter, and carriers densities. The sinusoidal and the cosinusoidal
like behaviors appear for respectively $\sigma^{\eta,0^\mathrm{o}}$ and
$\sigma^{\eta,90^\mathrm{o}}$ when varying the ellipticity angle $\varphi$
from 0 to $2\pi$. The strongest optical absorption of $\sigma^{\eta,0^\mathrm{o}}$
can be seen with the ellipticity angles at $\pi/2$ and $3\pi/2$ and the strongest absorption
along the perpendicular direction can be observed with the ellipticity angles at $0$ and
$2\pi$. There exists a phase displacement of $\pi/2$ when the sample is rotated
with $90^\mathrm{o}$ at a certain ellipticity angle. The strongest effect in the
anisotropy in the optical conductivities $\sigma_{xx}^\eta$ and $\sigma_{yy}^\eta$ can
be achieved at the ellipticity angles $\varphi=\pi/2$ and $3\pi/2$. We expect that such
an interesting feature in the ellipticity dependence of the optical absorption in a 2D SD system can
be observed experimentally through conventional infrared or THz transmission measurement
with elliptical polarizers.

In this study, we considered a 2D SD system with a device setting placed in
directions perpendicular to each other. Hereby, the anisotropic and tunable optical
absorption can be achieved by tuning the ellipticity ratio of the polarized radiation
light. As we know, a large difference in optical absorption can also be transformed
into electrical signals. As a result, 2D SD systems can be used in infrared
and THz optoelectronic devices. Moreover, we find that the effect of temperature, carrier
density and band-gap parameter can also modify significantly the optical and optoelectronic
properties of the 2D SD system. Due to the unique anisotropic and tunable optical
conductivity, the 2D SD system can be promising materials for applications in optics
and optoelectronics in infrared and THz bandwidths.

\section{Conclusions}
\label{sec:conclusions}

In this paper, we presented a detailed theoretical study on the anisotropic and
tunable optical conductivity of 2D SD systems in the presence of elliptically
polarized light irradiation. By considering the electron-photon interaction using
Fermi's gold rule, the optical conductivity was evaluated using the energy
balance equation approach derived from the semi-classical Boltzmann equation. The
effects of the ellipticity ratio of the polarized radiation field, temperature, carrier
density, and band-gap parameter on the optical conductivity was examined theoretically.
The main conclusions we have obtained from this study are summarized as follows.

2D SD systems exhibit anisotropic optical conductivity or optical
absorption within the bandwidth from THz to infrared in different geometrical
directions. This interesting phenomenon is induced by the anisotropy of the
electronic band structure of the system. In the presence of elliptically polarized
radiation field, the optical absorption can be effectively tuned by the ellipticity
ratio and the ratio of the optical conductivities along the longitudinal and vertical
directions $\sigma^{\eta,0^\mathrm{o}}(\omega)/\sigma^{\eta,90^\mathrm{o}}(\omega)$,
which increases with increasing ellipticity ratio of the polarized radiation fields. Therefore,
by changing the ellipticity ratio, the optical signals can be effectively detected
through a device setting based on the 2D SD materials. The ratio of $\sigma^{\eta,0^\mathrm{o}}(\omega)/\sigma^{\eta,90^\mathrm{o}}(\omega)$ increases
with temperature, which indicates that this effect can be very likely observed even at
room temperature. With increasing the carrier density, the optical absorption edge
is blue shifted, which is caused by the Pauli-blocking effects \cite{Grigorenko12}.
Using band-gap engineering \cite{Kim15,Chaves20}, the band structure of
the 2D SD system can be turned by changing the band-gap parameter resulting in a
tuning of the optical absorption in infrared and THz bandwidths.

From our theoretical results, we have found that
the optical and optoelectronic properties of the 2D SD systems can be effectively tuned
by the ellipticity ratio, temperature, carrier density, and band-gap parameter.
The 2D semi-Dirac systems can therefore be applied as tunable optical and optoelectronic
materials to be used for the realization of such devices as optical modulators,
switches, polarizers etc., which are active in the infrared to THz bandwidths.
Moreover, It has been shown that 2D SD electronic systems can be
obtained in materials such as multilayer (TiO$_2$)$_m$/(VO$_2$)$_n$
heterostructures \cite{Pardo09,Banerjee09}, $\alpha$-(BEDT-TTF)$_2$I$_{3}$ \cite{Kobayashi11,Suzumura13}, few-layer black phosphorus \cite{Makino17,Baik15}, silicene oxide \cite{Zhong17}, etc. By controlling the
ellipticity of THz or infrared polarized radiation with elliptical polarizers,
the elliptically dependence optical conductivity in this study can
be measured through the techniques such as Fourier transform
infrared (FTIR) spectroscopy and THz TDS measurement \cite{Kuzmenko08,Li08n,Arts18,Whelan20}.
Since the 2D SD materials mentioned above are usually predicted theoretically. It
is worth mentioning that our theoretical results would also be helpful for
detecting the 2D SD electrons in these materials.
We hope that our theoretical predictions can be verified experimentally in the near.

\section*{ACKNOWLEDGMENTS}
This work was supported by the National Natural Science foundation
of China (NSFC) (Grants No. 12004331, No. U1930116, No. U1832153,
No. U206720039, No. 11847054), Shenzhen Science and Technology
Program (No. KQTD20190929173954826), and by Yunnan Fundamental Research
Projects (No. 2019FD134). B.V.D. was supported through a post doc
fellowship from the Research Foundation Flanders (FWO-V1).


\begin{thebibliography}{99}

\bibitem{Novoselov04}
K. S. Novoselov, A. K. Geim, S. V. Morozov, D. Jiang, Y.
Zhang, S. V. Dubonos, I. V. Grigorieva, and A. A. Firsov,
\href{https://science.sciencemag.org/content/306/5696/666}
{Science \textbf{306}, 666 (2004)}.

\bibitem{Li17}
X. Li, L. Tao, Z. Chen, H. Fang, X. Li, X. Wang,
J.-B. Xu, and H. Zhu,
\href{http://dx.doi.org/10.1063/1.4983646}
{Appl. Phys. Rev. \textbf{4}, 021306 (2017)}.

\bibitem{Pardo09}
V. Pardo and W. E. Pickett,
\href{http://dx.doi.org/10.1103/PhysRevLett.102.166803}
{Phys. Rev. Lett. \textbf{102}, 166803 (2009)}.

\bibitem{Banerjee09}
S. Banerjee, R. R. P. Singh, V. Pardo, and W. E. Pickett,
\href{http://dx.doi.org/10.1103/PhysRevLett.103.016402}
{Phys. Rev. Lett. \textbf{103}, 016402 (2009)}.

\bibitem{Rodin14}
A. S. Rodin, A. Carvalho, and A. H. C. Neto,
\href{http://dx.doi.org/10.1103/PhysRevLett.112.176801}
{Phys. Rev. Lett. \textbf{112}, 176801 (2014)}.

\bibitem{Guan14}
J. Guan, Z. Zhu, and D. Tom\'{a}nek,
\href{http://dx.doi.org/10.1103/PhysRevLett.113.046804}
{Phys. Rev. Lett. \textbf{113}, 046804 (2014)}.

\bibitem{Rudenko15}
A. N. Rudenko, S. Yuan, and M. I. Katsnelson,
\href{http://dx.doi.org/10.1103/PhysRevB.92.085419}
{Phys. Rev. B \textbf{92}, 085419 (2015)}.

\bibitem{Dutreix16}
C. Dutreix, E. A. Stepanov, and M. I. Katsnelson,
\href{http://dx.doi.org/10.1103/PhysRevB.93.241404}
{Phys. Rev. B \textbf{93}, 241404(R) (2016)}.

\bibitem{Kobayashi11}
A. Kobayashi, Y. Suzumura, F. Pi\'{e}chon, and G. Montambaux,
\href{http://dx.doi.org/10.1103/PhysRevB.84.075450}
{Phys. Rev. B \textbf{84}, 075450 (2011)}.

\bibitem{Suzumura13}
Y. Suzumura, T. Morinari, and F. Pichon,
\href{http://dx.doi.org/10.7566/JPSJ.82.023708}
{J. Phys. Soc. Jpn. \textbf{82}, 023708 (2013)}.

\bibitem{Pyatkovskiy16}
P. K. Pyatkovskiy and T. Chakraborty,
\href{http://dx.doi.org/10.1103/PhysRevB.93.085145}
{Phys. Rev. B \textbf{93}, 085145 (2016)}.

\bibitem{Mawrie19R}
A. Mawrie and B. Muralidharan,
\href{http://dx.doi.org/10.1103/PhysRevB.100.081403}
{Phys. Rev. B \textbf{100}, 081403(R) (2019)}.

\bibitem{Chen18}
Q. Chen, L. Du, and G. A. Fiete,
\href{http://dx.doi.org/10.1103/PhysRevB.97.035422}
{Phys. Rev. B \textbf{97}, 035422 (2018)}.

\bibitem{Islam18}
SKFiroz Islam and A. Saha,
\href{http://dx.doi.org/10.1103/PhysRevB.98.235424}
{Phys. Rev. B \textbf{98}, 235424 (2018)}.

\bibitem{Mawrie19}
A. Mawrie and B. Muralidharan,
\href{http://dx.doi.org/10.1103/PhysRevB.99.075415}
{Phys. Rev. B \textbf{99}, 075415 (2019)}.

\bibitem{Carbotte19}
J. P. Carbotte, K. R. Bryenton, and E. J. Nicol,
\href{http://dx.doi.org/10.1103/PhysRevB.99.115406}
{Phys. Rev. B \textbf{99}, 115406 (2019)}.

\bibitem{Carbotte19s}
J. P. Carbotte and E. J. Nicol,
\href{http://dx.doi.org/10.1103/PhysRevB.100.035441}
{Phys. Rev. B \textbf{100}, 035441 (2019)}.

\bibitem{Sanderson18}
M. Sanderson, S. Huang, Y. Zhang, and C. Zhang,
\href{http://dx.doi.org/10.1088/1361-6463/aaba32}
{J. Phys. D: Appl. Phys. \textbf{51} 205302 (2018)}.

\bibitem{Dai19}
X. Dai, L. Liang, Q. Chen, and C. Zhang,
\href{http://dx.doi.org/10.1088/1361-648X/aafdd5}
{J. Phys.: Condens. Matter \textbf{31}, 135703 (2019)}.

\bibitem{Zhu21}
X. Zhu, W. Chen, X. Zhou, X. Xiao, and G. Zhou,
\href{http://dx.doi.org/10.1016/j.physe.2020.114462}
{Physica E \textbf{126}, 114462 (2021)}.

\bibitem{Sengupta21}
P. Sengupta and L. A. Jauregui,
\href{http://dx.doi.org/10.1063/5.0056116}
{J. Appl. Phys. \textbf{130}, 054303 (2021)}.

\bibitem{Saha16}
 K. Saha,
\href{http://dx.doi.org/10.1103/PhysRevB.94.081103}
{Phys. Rev. B \textbf{94}, 081103(R) (2016)}.

\bibitem{Dietl08}
P. Dietl, F. Pi\'{e}chon, and G. Montambaux,
\href{https://dx.doi.org/10.1103/PhysRevLett.100.236405}
{Phys. Rev. Lett. \textbf{100}, 236405 (2008).}

\bibitem{Sinha20}
P. Sinha, S. Murakami, and S. Basu,
\href{http://dx.doi.org/10.1103/PhysRevB.102.085416}
{Phys. Rev. B \textbf{102}, 085416 (2020)}.

\bibitem{Delplace10}
P. Delplace and G. Montambaux,
\href{https://journals.aps.org/prb/abstract/10.1103/PhysRevB.82.035438}
{Phys. Rev. B \textbf{82}, 035438 (2010).}

\bibitem{Saha17}
K. Saha, R. Nandkishore, and S. A. Parameswaran,
\href{http://dx.doi.org/10.1103/PhysRevB.96.045424}
{Phys. Rev. B \textbf{96}, 045424 (2017)}.

\bibitem{Makino17}
T. Makino, Y. Katagiri, C. Ohata, K. Nomura, and J. Haruyama,
\href{http://dx.doi.org/10.1039/C7RA03600K}
{RSC Adv. \textbf{7}, 23427 (2017)}.

\bibitem{Tajima09}
N. Tajima, S. Sugawara, R. Kato, Y. Nishio, and K. Kajita,
\href{http://dx.doi.org/10.1103/PhysRevLett.102.176403}
{Phys. Rev. Lett. \textbf{102}, 176403 (2009)}.

\bibitem{Banerjee12}
S. Banerjee and W. E. Pickett,
\href{https://journals.aps.org/prb/abstract/10.1103/PhysRevB.86.075124}
{Phys. Rev. B \textbf{86}, 075124 (2012)}.

\bibitem{Baik15}
S. S. Baik, K. S. Kim, Y. Yi, and H. J. Choi,
\href{http://dx.doi.org/10.1021/acs.nanolett.5b04106}
{Nano Lett. \textbf{15}(12), 7788-7793 (2015)}.

\bibitem{Kim15}
J. Kim, S. S. Baik, S. H. Ryu, Y. Sohn, S. Park,
B.g-G. Park, J. Denlinger, Y. Yi,
H. J. Choi, and K. S. Kim,
\href{http://dx.doi.org/10.1126/science.aaa6486}
{Science \textbf{349}(6249), 723-726 (2015)}.

\bibitem{Mak08}
K. F. Mak, M. Y. Sfeir, Y. Wu, C. H. Lui, J. A. Misewich,
and T. F. Heinz,
\href{http://dx.doi.org/10.1103/PhysRevLett.101.196405 }
{Phys. Rev. Lett. \textbf{101}, 196405 (2008)}.

\bibitem{Zhang18}
X. Zhang, W. Liao, H. Bao, and M. Zuo,
\href{http://dx.doi.org/10.1007/s00339-018-1776-1}
{Appl. Phys. A \textbf{124}:354 (2018)}.

\bibitem{Pedrotti17}
Frank L. Pedrotti, Leno M. Pedrotti, and Leno S. Pedrotti,
\emph{Introduction to Optics}
(Cambridge University Press, Cambridge, 2017).

\bibitem{Xu10}
W. Xu, H. M. Dong, L. L. Li, J. Q. Yao, P. Vasilopoulos,
and F. M. Peeters,
\href{http://dx.doi.org/10.1103/PhysRevB.82.125304}
{Phys. Rev. B \textbf{82}, 125304 (2010)}.

\bibitem{Bandurin16}
D. A. Bandurin, I. Torre, R. K. Kumar, M. B. Shalom, A.
Tomadin, A. Principi, G. H. Auton, E. Khestanova, K. S.
Novoselov, and I. V. Grigorieva,
\href{http://dx.doi.org/10.1126/science.aad0201}
{Science \textbf{351}, 1055 (2016)}.

\bibitem{Sun08}
D. Sun, Z.-K. Wu, C. Divin, X. Li, C. Berger,W. A. de Heer,
P. N. First, and T. B. Norris,
\href{http://dx.doi.org/10.1103/PhysRevLett.101.157402}
{Phys. Rev. Lett. \textbf{101}, 157402 (2008)}.

\bibitem{Samal21}
S. S. Samal, S. Nandy, and K. Saha,
\href{http://dx.doi.org/10.1103/PhysRevB.103.L201202}
{Phys. Rev. B \textbf{103}, L201202 (2021)}.

\bibitem{Xiao16}
Y. M. Xiao, W. Xu, B. Van Duppen, and F. M. Peeters,
\href{http://dx.doi.org/10.1103/PhysRevB.94.155432}
{Phys. Rev. B \textbf{94}, 155432 (2016)}.

\bibitem{Grigorenko12}
A. N. Grigorenko, M. Polini, and K. S. Novoselov,
\href{http://dx.doi.org/10.1038/nphoton.2012.262}
{Nat. Photon. \textbf{6}, 749 (2012)}.

\bibitem{Kim21}
H. Kim, S. Z. Uddin, D.-H. Lien, M. Yeh,
N. S. Azar, S. Balendhran, T. Kim, N. Gupta,
Y. Rho, C. P. Grigoropoulos, K. B. Crozier, and A. Javey,
\href{http://dx.doi.org/10.1038/s41586-021-03701-1}
{Nature (London) \textbf{596}, 232 (2021)}.

\bibitem{Chaves20}
A. Chaves, J. G. Azadani, H. Alsalman, D. R. da Costa, R. Frisenda,
A. J. Chaves, S. H. Song, Y. D. Kim, D. He, J. Zhou et al.,
\href{http://dx.doi.org/10.1038/s41699-020-00162-4}
{npj 2D Mater. Appl. \textbf{4}, 29 (2020)}.

\bibitem{Zhong17}
C. Zhong, Y. Chen, Y. Xie, Y.-Y. Sun, and S. Zhang,
\href{http://dx.doi.org/10.1039/c6cp08439g}
{Phys. Chem. Chem. Phys. \textbf{19}, 3820 (2017)}.

\bibitem{Kuzmenko08}
A. B. Kuzmenko, E. van Heumen, F. Carbone, and D. van der Marel,
\href{http://dx.doi.org/10.1103/PhysRevLett.100.117401}
{Phys. Rev. Lett. \textbf{100}, 117401 (2008)}.

\bibitem{Li08n}
Z. Q. Li, E. A. Henriksen, Z. Jiang, Z. Hao, M. C. Martin,
P. Kim, H. L. Stormer, and D. N. Basov,
\href{http://dx.doi.org/10.1038/nphys989}
{Nature Phys. \textbf{4}, 532 (2008)}.

\bibitem{Arts18}
K. Arts, R. Vervuurt,  A. Bhattacharya, J. G. Rivas,
J. W. Oosterbeek, and A. A. Bol,
\href{http://dx.doi.org/10.1063/1.5044265}
{J. Appl. Phys. \textbf{124}, 073105 (2018)}.

\bibitem{Whelan20}
P. R Whelan, Q. Shen, B. Zhou, I G Serrano, M V. Kamalakar,
D. M A Mackenzie, J. Ji, D. Huang, H. Shi, D. Luo et al.,
\href{http://dx.doi.org/10.1088/2053-1583/ab81b0}
{2D Mater. \textbf{7}, 035009 (2020)}.

\end{thebibliography}
\end{document}